\documentstyle[sprocl,epsfig]{article}
\def\be{\begin{equation}}
\def\ee{\end{equation}}
\def\beq{\begin{eqnarray}}
\def\eeq{\end{eqnarray}}

\begin{document}

\begin{flushright}
Preprint WUB 97-18 \\
hep-ph/9706224
\end{flushright}

\vspace{1em}

\title{PREDICTIONS FOR THE $\eta_c\gamma$ TRANSITION FORM FACTOR}

\author{\underline{Thorsten Feldmann} and Peter Kroll}

\address{Department of Theoretical Physics, University of Wuppertal,\\
    D-42097 Wuppertal, Germany}


\maketitle\abstracts{
The $\eta_c\gamma$ transition form factor is calculated
in a model based on a modified hard scattering approach to 
exclusive reactions, in which transverse degrees of
freedom are taken into account.
For the $\eta_c$-meson a distribution amplitude of the
Bauer-Stech-Wirbel type is used, where
the two free parameters, namely the decay constant
$f_{\eta_c}$ and the transverse size of the wave function,
are related to the Fock state probability and the
width for the two-photon decay $\Gamma_{\eta_c \to \gamma\gamma}$.
}

\begin{center}
{\it Talk presented at the conference {\sc Photon '97},\\ Egmond aan Zee,
The Netherlands, May 10-15, 1997}
\end{center}

\section{Introduction}

At large momentum transfer the hard scattering approach
(HSA)~\cite{BrLe80} provides a scheme to calculate exclusive
processes. Observables are described as convolutions
of hadronic wave functions, which embody soft non-perturbative
physics, and hard scattering amplitudes to be calculated
from perturbative QCD.

One interesting class of such observables 
are the meson-photon transition form factors,
which are at leading order of
purely electromagnetic origin. Hence, the uncertainties related to
the appropriate value of the strong coupling constant or
the size of the (Feynman) contributions coming from the overlap of the soft
wave functions are absent.
For example, in case of the pion-photon transition 
form factor it has been shown recently
that the experimental data can be
well described by a distribution amplitude that is close
to the asymptotic one;~\cite{KrRa96} and for the
$\eta$ and $\eta'$ mesons a determination of the decay
constants and the mixing angle from the measurement of
their transition form factors is possible.~\cite{Jakob:1996}

Here we discuss the application of the HSA to the $\eta_c\gamma$ transition
form factor.
In this case the finite mass of the charmed quarks always
provides a large scale which allows
the application of the HSA even for zero virtuality of the
probing photon, $Q^2\to 0$.
Then 
the HSA result for the transition form factor at $Q^2=0$ can be
related to 
the decay width $\Gamma[\eta_c \to
\gamma\gamma]$, 
whereas the shape turns out to be unique in the $Q^2$ region of
experimental interest and for reasonable values of the
valence Fock state probability $P_{c\bar c}$.

\section{Modified Hard Scattering approach}

The $\eta_c\gamma$ transition form factor in the
modified HSA~\cite{BrLe80,bot:89} is defined in analogy to
the $\pi\gamma$ form factor~\cite{KrRa96} in terms of a
hard scattering amplitude $T_H$, a non-perturbative (light-cone)
wave function $\Psi_0$ of the leading $|c\bar c\rangle$ Fock state
and a Sudakov factor as
\beq
F_{\eta_c\gamma} (Q^2)
        &=&
\int_0^1 dx  \int \frac{d^2 {\vec b_\perp}}{4\pi} 
        \hat \Psi_0(x, {\vec b_\perp})  \hat T_H(x,{\vec b_\perp},Q) 
        \exp\left[ - {\mathcal S}(x,{\vec b_\perp}, Q) \right]
\label{Fdef}
\eeq
Here $\vec b_\perp$ denotes the transverse size in configuration
space, and $x$ is the usual Feynman parameter.
In the present case, 
the Sudakov factor  $\exp[ - {\mathcal S}]$ 
can be neglected for two reasons:
First, due to the large  quark mass the radiative
corrections only produce soft divergences but no collinear ones.
Secondly, in contrast to the
light meson case where the Sudakov factor provides a consistent
tool to suppress
the contributions from the endpoint regions where perturbation
theory becomes unreliable, 
the distribution amplitude $\phi(x)$ 
in the $\eta_c$ meson
is expected to be strongly peaked at $x = 1/2$, and the potentially dangerous
endpoint regions are unimportant anyway. 
It is then more appropriate to use the
Fourier transformed definition of the form factor,
\beq
F_{\eta_c\gamma} (Q^2)
        &\simeq&
\int_0^1 dx \, \int \frac{d^2 {\vec k_\perp}}{16 \pi^3} \,
        \Psi_0(x, {\vec k_\perp}) \, T_H(x,{\vec k_\perp},Q) 
\label{Fdef2}
\eeq
The hard scattering amplitude in leading order is up to 
conventional normalization constants calculated from
the following Feynman diagrams ($\bar x=(1-x)$).

\unitlength1cm
\begin{center}
\epsfclipon
{\psfig{file=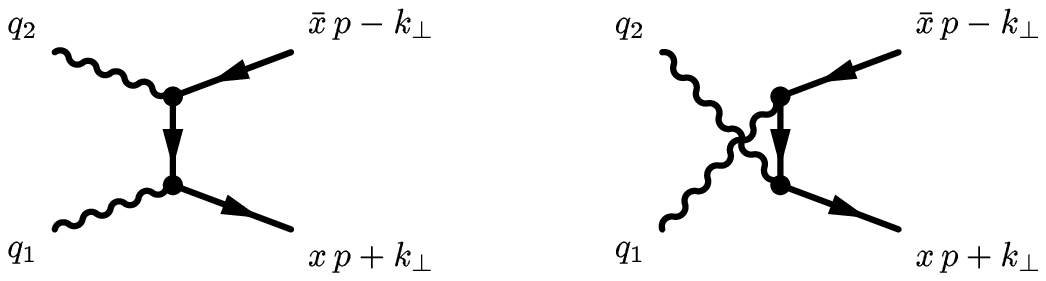, bb = 135 650 445 735,  width = 7.5cm}}
\end{center}
With one photon being almost on-shell $q_1^2 \simeq 0$ and
the virtuality of the second photon denoted as $q_2^2 = -Q^2$, this
leads to
\beq
T_H(x,{\vec k_\perp},Q) &=& e_c^2 \, 2 \sqrt 6  \ 
\frac{2}{x\, Q^2 + ( x \bar x + \rho^2) \, M^2 + {\vec k_\perp^2}}
+ O(\alpha_s)
\label{THhat2}
\eeq
Here $p^2 = M^2$ and $\rho := m_c/M \simeq 0.5$, and $e_c$ denotes
the fractional charge of the charm quark.

For the wave function, it is reasonable to assume the following
factorizing form 
\beq
\Psi_0(x, {\vec k_\perp}) = \frac{f_{\eta_c}}{2 \sqrt{6}} \,
\phi(x) \, \Sigma({\vec k_\perp})
\eeq
Here $f_{\eta_c}$ is the decay constant (corresponding to
$f_\pi = 133$~MeV), and $\phi(x)$ is
the quark distribution amplitude (DA) in the meson.
In the following we will use
a form of the wave function
adapted from
Bauer, Stech and Wirbel~\cite{BSW85}
\beq
 \phi(x) &=& N_\phi(a) \ x \, \bar x \
 \exp\left[ - a^2 \, M^2 \, \left(x - 1/2\right)^2\right]
\eeq
The normalization constant $N_\phi(a)$ is determined from
$\int_0^1 dx \, \phi(x)~=~1$.
Note that this DA is concentrated around $x=1/2$.
Furthermore, $\Sigma$
is a Gaussian shape function which takes into
account the finite transverse size of a meson 
~\footnote{In fact, we
use the same size parameter $a$ for both, the distribution
amplitude and the transverse shape. Strictly speaking, this
equality only holds in the non-relativistic limit.}
\beq
\Sigma({\vec k_\perp}) 
  &=& 16 \pi^2 \, a^2 \, e^{-a^2 \, {\vec k_\perp^2}}
\label{Sigmafunc}
\eeq

\section{Fixing the parameters}

The parameters entering
the wave function are
constrained by
the Fock state probability
\beq
 1 \geq P_{c\bar c} &=& 
 \int \frac{dx \, d^2{\vec k_\perp}}{16 \pi^3} \,
 \left| \Psi_0(x,{\vec k_\perp}) \right|^2
\eeq
One expects $0.8 \leq P_{c\bar c} < 1$, and we find
that for a given value of $f_{\eta_c}$ the form factor only
mildly depends on the value of $P_{c\bar c}$, such that
we may use $P_{c\bar c}=0.8$ as a constraint for the size
parameter $a$ which leads to $a \simeq 1$~GeV$^{-1}$,
in consistence with typical estimates for the radius
$\langle r^2 \rangle = 3 \,a^2 \simeq (0.4~{\rm fm})^2$ or the 
quark velocity $v^2 =3/(Ma)^2 \simeq 0.3$.

For such values of $a$ and $M$ it makes sense to first
consider the collinear limit $(aM)^2 \gg 1$ such that the
wave function collapses to $\delta$~distributions around
$x=1/2$ and $\vec k_\perp=0$. Accordingly,
\beq
F_{\eta_c\gamma}(Q^2) 
&  = & 
e_c^2 \, f_{\eta_c} \, \frac{4}{M^2 + Q^2 }
\, \left(1 + O(1/a^2) + O(\alpha_s) \right)
\label{hsalimit}
\eeq
Note that this structure of the form factor is similar to the
vector meson dominance prediction.

The form factor at $Q^2=0$ is related to the decay rate
$\Gamma[\eta_c\to\gamma\gamma]$ which still suffers
from large experimental uncertainties~\cite{PDG96}
\beq
&& \Gamma_{\eta_c \to \gamma\gamma} =
\frac{e^4 M^3}{64 \pi} 
  \, \left| F_{\eta_c\gamma}(0) \right|^2  \
=\left\{\begin{array}{l}
 7.5^{+1.6}_{-1.4}~\mbox{keV} \quad \mbox{(direct)} \\  
(4.0 \pm 1.5~\mbox{keV}) \cdot  \frac{\Gamma^{\rm
    tot}_{\eta_c}}{13.2~{\rm MeV}} \quad \mbox{(BR)}
\end{array}  \right. 
\eeq
In the non-relativistic limit
this can also be related to the partial width for
$J/\Psi \to e^+e^-$ with $f_{\eta_c}\simeq f_{J/\Psi}
\simeq 400$~MeV. However the $\alpha_s$ corrections are
known to be large,~\cite{Ba79} and the relativistic corrections are
large and model dependent. Typically one finds~\cite{AhMe95,HwKi97,Hu96}
 $f_{\eta_c}/f_{J/\Psi}
=1.2\pm0.1$ and $\Gamma_{\eta_c\to \gamma\gamma} = (5-7)$~keV.
In the following, we will therefore use $\Gamma_{\eta_c\to\gamma\gamma}$
as a physical normalization for the form factor.
In the region of experimental
interest, $Q^2 \leq M^2$, we assume that the
additional $Q^2$ dependence of the form factor which is induced
by perturbative QCD corrections 
is small. For larger $Q^2$ the  evolution of
the wave function~\cite{BrLe80} and the $\alpha_s$ corrections to
the hard scattering amplitude~\cite{Br83} will become
important.

\section{Results}

\begin{figure}[hbtp]
\unitlength1cm
\begin{center}
{\psfig{file=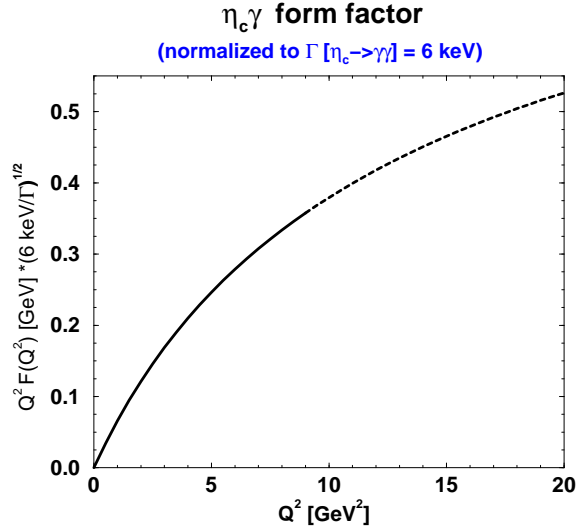, width=7cm, angle = -90}}
\end{center}
\caption{The $\eta_c\gamma$ 
 form factor rescaled by $Q^2$ and normalized to
 $\Gamma_{\eta_c\to\gamma\gamma} = 6$~keV in leading order of the HSA.
For $Q^2>M^2$ (dashed line) the perturbative QCD
corrections 
are expected to become important.}
\label{fig}
\end{figure}

Fig.~\ref{fig} shows the scaled form factor $Q^2 \, F_{\eta_c\gamma}$
normalized to 
$\Gamma_{\eta_c\to\gamma\gamma} = 6$~keV.
The following conclusions can be drawn:
The shape of the form factor predicted
      by the HSA approach 
      is unique. It can be well approximated
      by \beq F_{\eta_c\gamma}(Q^2) &\simeq&
      \frac{F(0)}{1 + Q^2/(M^2 + 2\, \langle \vec k_\perp^2 \rangle)}
      \label{interpol} \eeq 
      which takes into account the leading corrections to
      the collinear approximation, reducing
      the form factor at $Q^2=0$ by order 10\%. In our case we have
      $M^2 + 2\,\langle \vec k_\perp^2 \rangle \simeq (3.2~{\rm GeV})^2$
      which is not much larger than the value for the $J/\Psi$ mass
      that one would have inserted in the VDM approach.
      Note that eq.~(\ref{interpol}) may be of particular use
      for the analysis of the decay width $\Gamma_{\eta_c\to
      \gamma\gamma}$.

The decay constant $f_{\eta_c}$ enters the form factor
      as an overall factor. Thus, in principle one may
      determine its value from a precise measurement of
      $F_{\eta_c\gamma}$ and/or $\Gamma_{\eta_c \to \gamma\gamma}$.
      For this purpose also the perturbative corrections to the
      form factor at arbitrary $Q^2$ should be taken into account
      in a consistent way, which is to be analyzed in a forthcoming
      paper. 
      In this context more precise information
      from other theoretical approaches (lattice, QCD sum rules)
      is of course welcome.

\section*{Acknowledgments}
T.F.\ is supported by
the {\it Deutsche Forschungsgemeinschaft}.

\section*{References}
\bibliography{ref}

\end{document}